\documentclass{ws-ijmpe}

\usepackage{amsmath,multirow,multicol}

\begin{document}

\markboth{T. V. Nhan Hao, J. Le Bloas, Meng-Hock Koh, L. Bonneau,
P. Quentin}{Further microscopic studies of the fission barriers of
  heavy nuclei}

\catchline{}{}{}{}{}

\title{Further microscopic studies
of the fission barriers of heavy nuclei}

\author{T. V. NHAN HAO}

\address{Tan Tao University, Department of Engineering, Tan Tao
  University avenue, Tan Duc Ecity, Long An Province,
  Vietnam;\\
Univ. Bordeaux, CENBG, UMR5797, 33170 Gradignan, France; \\
CNRS, IN2P3, CENBG, UMR5797, CENBG, 33170 Gradignan, France.}

\author{J. LE BLOAS}

\address{Univ. Bordeaux, CENBG, UMR5797,
33170 Gradignan, France; \\
CNRS, IN2P3, CENBG, UMR5797, CENBG, 33170 Gradignan, France.}

\author{MENG-HOCK KOH}

\address{Universiti Teknologi Malaysia, 81310 Skudai, Johor, Malaysia; \\
Univ. Bordeaux, CENBG, UMR5797, 33170 Gradignan, France; \\
CNRS, IN2P3, CENBG, UMR5797, CENBG, 33170 Gradignan, France.}

\author{L. BONNEAU, P. QUENTIN\footnote{Corresponding
    author: quentin@cenbg.in2p3.fr}}

\address{
Univ. Bordeaux, CENBG, UMR5797, 33170 Gradignan, France; \\
CNRS, IN2P3, CENBG, UMR5797, CENBG, 33170 Gradignan, France.}



\maketitle

\begin{history}
\received{(received date)}
\revised{(revised date)}
\end{history}

\begin{abstract}

Two systematic sources of error in most current microscopic
evaluations of fission-barrier heights are studied. They are concerned
with an approximate treatment of the Coulomb exchange terms (known as
the Slater approximation) in the self-consistent mean fields and the
projection on good parity states (e.g., of positive parity for the
spontaneous fission of an even-even nucleus) of left-right reflection
asymmetric intrinsic solutions (e.g., around the second
barrier). Approximate or unprojected solutions are shown to lead each
to an underestimation of the barrier heights by a few hundred keV.

\end{abstract}

%
%

\section{Introduction}

In most microscopic calculations of nuclear binding energies using
effective nucleon-nucleon interaction or their Energy Density
Functional (EDF) avatars, data are reproduced in the best cases
with an accuracy which hardly goes below a couple of MeV. When evaluating
relative energies, one may hope for a partial cancellation of such
errors but this is by far not granted in general. A particular case of
choice of such relative energies is constituted by the so-called
fission barrier heights. Being a rough one-dimensional description of
the very complex many-body fission process, their "experimental"
values result in most cases from a model dependent -or partial-
analysis of some data. Nevertheless, they provide undoubtedly a much
wanted energy scale on which microscopic calculations adjust their
phenomenological parameters (e.g. this has been in particular the case
for the fit of the SkM$^\ast$ Skyrme interaction~\cite{SKM}). As an
example of the accuracy of these barrier calculations we may quote the
results of Ref.~\cite{BQS} where an average accuracy of about 1 MeV
for the first and second barriers of actinide nuclei (as well as
fission isomer excitation energies) is found upon making a very rough
estimate of the rotational energy correction.

Such a range of error is very significant in physical terms. It is
generally estimated that a variation of 1 MeV roughly corresponds to a
change of spontaneous fission half lives by 4 orders of magnitude. It
is therefore of paramount importance to pinpoint systematical errors
inherent to a given theoretical (phenomenological) approach which
could lead to wrong estimates of fission properties or conversely, when
making a fit of the interaction taking care of fission barriers,
which could have adverse effects on
theoretical estimates of other quantities.

Without aiming at any exhaustibility in chasing such errors, two
examples will be studied here: i) the approximate treatment of the
Coulomb exchange terms known as the Slater approximation in the
self-consistent mean fields, ii) the projection on good parity states
(e.g. of positive parity for the spontaneous fission of an even-even
nucleus) of left-right reflection asymmetric intrinsic solutions
(e.g. around the second barrier).

%
%

\section{On the effect of the Slater approximation of Coulomb exchange
  terms}

Most microscopic calculations use the infinite matter Pauli
correlation function (within a local density approximation) to compute
the Coulomb exchange expectation value in a Slater determinant
producing an EDF piece involving only the local proton density
\begin{equation}
E^{\tiny{\rm Slater}}_{\tiny{\rm Coul.exch.}} =
-\frac{3}{4}e^2{\left(\frac{3}{\pi}\right)}^{\frac{1}{3}}\int d^3r \:
\rho_{p}^{4/3}(\vec{r}) \:.
\end{equation}
The above yields, within a variational approach, the following mean
field piece
\begin{equation}
V^{\tiny{\rm Slater}}_{\tiny{\rm Coul.exch.}}(\vec{r}) =
-{\left(\frac{3}{\pi}\right)}^{\frac{1}{3}}e^2\rho_{p}^{1/3}(\vec{r}) \:.
\end{equation}

This approach is usually refered to as the Slater
approximation~\cite{Slater}. It has been tested many years
ago~\cite{TSQ} by incorporating an exact Coulomb exchange treatment in
Hartree-Fock calculations of the ground state (spherical or deformed)
solutions of 8 light nuclei (from $^{16}$O to $^{56}$Ni). In this early work,
the Skyrme SIII interaction~\cite{SIII} had been used.

Such a study has been recently revisited by J. Skalski~\cite {Skal} for
the ground-state Hartree-Fock solutions of  9 spherical subshell closed
nuclei, from nuclei as light as $^{16}$O to superheavies such as
$^{310}$126. The SkP~\cite{SkP} and SkM$^\ast$ Skyrme interactions
have been used.

From these two studies, one has concluded that the errors found
(namely relative errors, consistently in all this Section) for the
Coulomb exchange energies, are relatively small, amazingly
interaction-independent (at least for the three interactions under
consideration, namely SIII, SkM$^\ast$ and SkP) and decreasing with the
total particle number $A$ (as expected in view of the infinite matter
origin of the Slater approximation).

Even though no quantitative account nor general argument, has been given there,
as far as fission
barrier heights are concerned, one should mention the calculations
of Ref.~\cite{ANG}. The authors have included in a HFB framework full
Coulomb exchange terms in various nuclear systems , including the fission
barriers of $^{254}$No.

In the following we have performed self-consistent (Hartree-Fock or
Hartree-Fock-plus-BCS calculations) using only the SIII Skyrme
effective interaction (with a seniority interaction for the
BCS part). The Coulomb energy and mean-field contributions have been
calculated as proposed in Ref.~\cite{PhQ} by evaluating the Coulomb
matrix elements upon using a  gaussian integral representation of a
Yukawa-type two-body interaction as
\begin{equation}
\frac{e^{-\mu|\vec{r}_1-\vec{r}_2|}}{|\vec{r}_1-\vec{r}_2|} =
\sqrt{\frac{2}{\pi}}\int^{\infty}_0e^{-\mu^2\sigma^2/2}
e^{-\mu|\vec{r}_1-\vec{r}_2|^2/2\sigma^2}\frac{d\sigma}{\sigma^2} \:.
\end{equation}

First, we have extended our study of the above spherical subshell
closed nuclei beyond the Hartree-Fock approximation in use and allowed
for pairing correlations \`a la BCS with a seniority force. For two
nuclei ($^{90}$Zr and $^{298}$114) where pair correlated solutions
have been found to occur, the error has been dramatically reduced (by much more
than half as seen in Fig.~1). This appears thus as a hint that the
error found so far is due to the particular filling of those nuclei
which had been considered.
\begin{figure}[h]
\begin{center}
\includegraphics[width=0.75\textwidth]{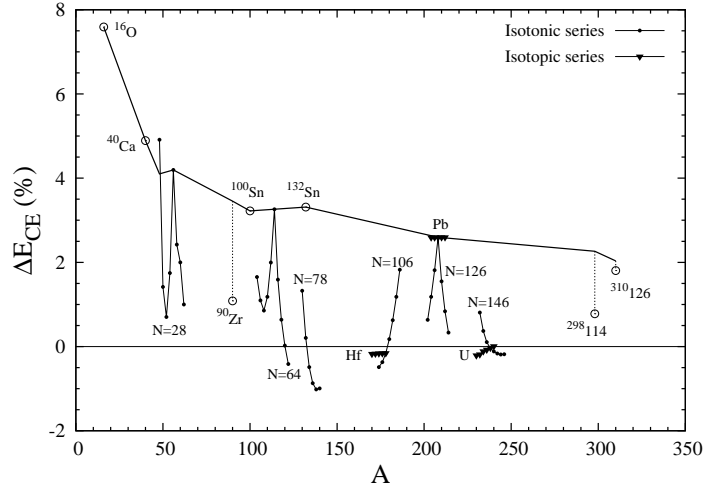}
\caption{Relative error $\Delta E_{\rm CE}$ associated with open
  proton shell nuclei, where $E_{\rm CE}$ is the expectation value of
the exchange part of the Coulomb interaction. The closed proton shell nuclei are reported as
  open circles linked by the solid line. Some isotonic series are
  represented in dash lines, whereas three isotopic series
  corresponding to Hf, Pb, and U isotopes are plotted with black
  triangles. For the $^{90}$Zr and $^{298}$114 nuclei, the plotted solutions
correspond to BCS solutions.}
\end{center}
\end{figure}

We have then calculated the error for six isotonic series ($N =28$, 64,
78, 106, 126, 146) of nuclei constrained to be spherically symmetric
(allowing pairing correlations to be included). As seen in Fig.~1 the
trend of the error is totally unambiguous. Away from proton magic
numbers ($Z = 20$, 28, 50, 82) the quality of the Slater approximation
improves very rapidly. To confirm the role of the proton single-particle
level scheme, we have performed similar calculations for three
short isotopic series ($Z = 72$, 82, 92). The error is found, see
Fig.~1, to be remarkably stable within a given isotopic series.

Then, we have explored changes in the proton single-particle level density
$\overline{{\rho}_p}$ at the Fermi surface, not upon changing the proton
number but instead the axial quadrupole deformation for the same
nucleus. As examplified in Fig.~2 reporting the results of calculations
for 5 nuclei ($^{24}$Mg, $^{48}$Cr, $^{106}$Mo, $^{178}$Hf, $^{238}$U)
the error is always--and often very significantly--larger for their
deformed ground-state solutions than for the solutions constrained to
have a vanishing quadrupole moment.
\begin{figure}[h]
\begin{center}
\includegraphics[width=0.75\textwidth]{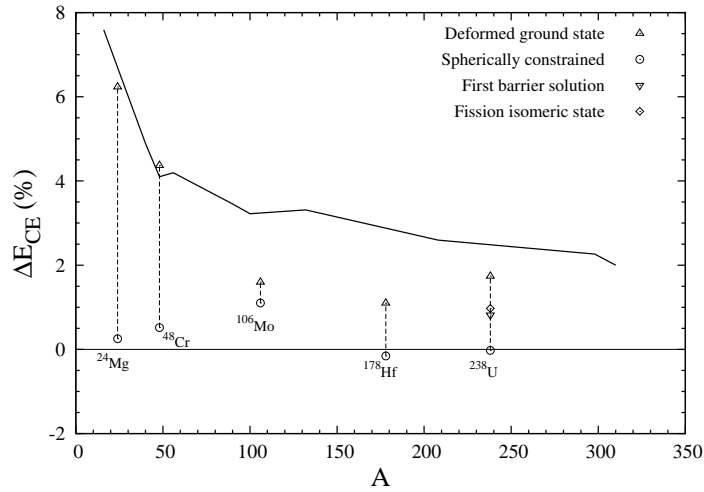}
\caption{Relative errors $\Delta E_{\rm CE}$ in the ground-state
  solution of deformed nuclei and in their spherically constrained
  solution. The solid line, shown for comparison, corresponds to the
  closed proton shell nuclei appearing in Fig.~1.}
\end{center}
\end{figure}
In the case of $^{238}$U, we
have also reported the errors corresponding to the top of the first
barrier and the fission isomeric state. These errors lie between those
obtained for the spherical and deformed solutions. To
confirm these findings we have compared the error made on the Coulomb exchange
energies (middle panel of Fig.~3) with the proton BCS pairing
gap, taken as an index of the proton single-particle level density at
the Fermi surface (lower panel of Fig.~3) along the beginning of
the calculated fission barrier of the $^{70}$Se nucleus imposing a
vanishing axial octupole moment (upper panel of
Fig.~3).
\begin{figure}[h]
\begin{center}
\includegraphics[width=0.5\textwidth]{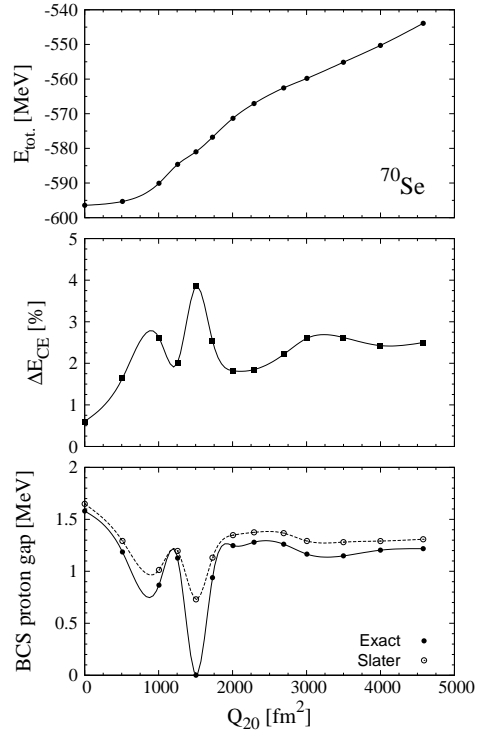}
\caption{(Top panel) Deformation-energy curve of the $^{70}$Se as a
function of the mass quadrupole moment $Q_{20}$ in $\rm
fm^22$. (Middle panel) Relative error $\Delta E_{\rm CE}$ as a
function of $Q_{20}$. (Bottom panel) Variation with $Q_{20}$ of proton
pairing gaps calculated with the exact treatment of Coulomb exchange
terms in solid line (with full circles) and using the Slater
approximation in dashed line (with open circles).
}
\end{center}
\end{figure}
As seen from Fig.~3 the error and the gap are very well
anti-correlated functions of the deformation.

We can therefore conclude that the quality of the Slater approximation
will be less good for low values of $\overline{{\rho}_p}$ than for
densities corresponding to mid-shell situations.

Since it is well known that systematically $\bar{\rho}$
will be higher at the top of the fission barrier than at the ground
state deformation (for a direct evidence of that fact in the
context of fission-barrier calculations within the Hartree-Fock-plus-BCS
framework, see Ref.~\cite{BQ}),
we can predict that the correction to be made to
account for the error brought in by the Slater approximation will be
lower in the former case than in the latter. As a result,
using the Slater approximation will systematically lead to an
underestimation of the fission-barrier heights. In the calculations
already discussed (see Fig.~2) of the $^{238}$U fission
barrier~\cite{JLBKMH}, one should add beyond the Slater approximation
values about 310 keV to the first fission barrier height and about 280
keV to the fission isomeric energy, quantities which are not without
consequences on spontaneous fission halflives as already mentioned.
Even though it is hard
to get more than a tendency on the l.h.s part of the Figure 12 of Ref.~\cite{ANG}
their results (for $^{254}$No) are consistent with our above conclusions.

Before closing this Section, two facts deserve a short notice.

In the above discussion of the impact of $\overline{{\rho}_p}$ on the
quality of the Slater approximation, we have only considered the
errors on the Coulomb exchange energies. Of course, the quantities
whose errors should have been
compared (and actually which have been discussed just above in the
$^{238}$U fission study) are total energies (the sum of kinetic, nuclear,
direct and exchange Coulomb energies). However, it has been
demonstrated~\cite{JLBKMH} that, in most cases, the errors on such
total energies and Coulomb exchange energies are equal up to some tens
of keV.

As found out already in the results of Ref.~\cite{TSQ} and emphasized
in Ref.~\cite{Skal}, it has been systematically found that all hole
states, in Hartree-Fock parlance, are pushed up by the Slater
approximation from their location in exact calculations, while conversely, all
particle states are pushed down. As a result the proton single
particle level density at the Fermi surface is unduly enhanced by the
Slater approximation. The effect of such an error on proton
particle-hole energies as far as
correlation properties are concerned, remain to be
specifically studied. Indeed, if the residual interaction is to be
considered as a perturbation, it is easy to see that such a trend
could affect significantly the configuration mixing.

%
%

\section{On the effect of a projection on good parity states}

For a very long time, it is known~\cite{MN} that the second fission
barriers of actinides are unstable with respect to the left-right
asymmetry. It has been then recognized that this phenomenon is driven
by the heavy fragment shell effects explaining thus already at this
early stage of the fission process, the asymmetric pattern of the
fragment mass yields (at low compound-nucleus energy) which had been
observed in this region long before.

In most microscopic or macro-microscopic
calculations in the U, Pu region, upon increasing the elongation after
the fission isomeric state, the intrinsic equilibrium solution becomes
unstable with respect to the left-right reflection symmetry and acquires rapidly
a larger and larger octupole deformation which stabilizes at a value
corresponding, as we just have noted, to the most probable
fragmentation. This fact, as we will see, plays an important role for
our purpose here.

Yet, even though the intrinsic parity may be broken for some
microscopic solutions under consideration, the parity of the physical
solution must be conserved during the fission process. For instance,
if one describes the spontaneous fission of an even-even nucleus as
$^{240}$Pu, one should evaluate the fission barrier obtained upon
imposing the positive parity to the solutions. This may be obtained by
projecting intrinsic solutions on the desired parity.
This corresponds thus to a projection after variation,
amounting to a mixing of two configurations $|\Psi\rangle$ and
$|\widetilde{\Psi}\rangle = \hat{\Pi}|\Psi\rangle$, where $\hat{\Pi}$
is the parity operator.

Recently such a projection of correlated microscopic solutions \`a la
HTDA (Highly Truncated Diagonalization Approach)~\cite{PQL} has been extensively studied as the PhD thesis work of
one of the authors~\cite{Hao}. We will thus simply outline here, the major
points of this approach and discuss the results which have been
obtained when applying it to the second fission positive-parity
barrier of $^{240}$Pu.

The HTDA calculations are designed to
produce realistic correlated wave functions in an approach which
conserve explicitly the particle number and do not violate the
Pauli principle. They consist in performing intrinsic shell-model like
calculations using single-particle states deduced from a mean-field potential
$\hat{U}$. This field is a priori arbitrary. Yet, its realistic
character, given a state-of-the-art microscopic effective Hamiltonian
$\hat{H}$, is essential to limit the size of our $n$-particle $n$-hole
many-body basis. The mean field $\hat{U}$ will be taken here as the
one-body reduction of the nucleon-nucleon effective interaction
$\hat{V}$ included in $\hat{H}$ for a corresponding approximate
(e.g., \`a la BCS) microscopic correlated wave function.

The HTDA Hamiltonian is written as
\begin{equation}
\hat{H}=\hat{K}+\hat{V}=\hat{H}_{\mathrm{MF}}+\hat{V}_{\mathrm{res}}
\end{equation}
where $\hat{K}$ is the kinetic energy while the mean field (one body)
Hamiltonian $\hat{H}_{\mathrm{MF}}$ is defined by
\begin{equation}
\hat{H}_{\mathrm{MF}}=\hat{K}+\hat{U}-C
\end{equation}
and the residual interaction $\hat{V}_{\mathrm{res}}$ by
\begin{equation}
\hat{V}_{\mathrm{res}}=\hat{V}-\hat{U}+C
\end{equation}

The particle-hole vacuum $|{\Phi}_0\rangle$ is defined as an
eigensolution of $\hat{H}_{\mathrm{MF}}$. Here we have chosen this
vacuum as the ground state of $\hat{H}_{\mathrm{MF}}$. The constant
$C$ appearing in the above definitions is chosen in such a way that
the expectation value of the residual interaction is vanishing for
$|{\Phi}_0\rangle$.

In the present calculations, we have restricted the particle and hole
single-particle states to lie within 6 Mev below and above the Fermi
energy (defined as half the sum of the energies of the last occupied
and first unoccupied state in $|{\Phi}_0\rangle$). The many body basis
is comprised, beyond the vacuum, of only one-pair-transfer states (pairs
meaning here Cooper pairs of time-reversal conjugate single-particle
states). We have chosen the Skyrme SkM$^\ast$ for $\hat{V}$. As usually
done in practical HTDA calculations, we have replaced $\hat{V}$ by a
density-independent zero-range interaction to define the residual
interaction. The correlated wave function $|\Psi\rangle$ is obtained
by a diagonalization of the Hamiltonian $\hat{H}$ through a Lancz\"os
algorithm process.

Through a judicious choice of the mean field $\hat{U}$ (see
Ref.~\cite{Hao}) the expression for the energy of the projection of
the state $|\Psi\rangle$ on a state of good parity $p = \pm 1$ may
be simplified as
\begin{equation}
E_{p}=\frac{\langle \Psi|\hat{H}|\Psi\rangle + p \langle
\Psi|\hat{H} |\widetilde{\Psi}\rangle}{1+p\langle
\Psi|\widetilde{\Psi}\rangle} \:.
\end{equation}
The non-diagonal overlaps are calculated using methods due to
L\"owdin~\cite{Lov} (The use of the heavy Balian-Br\'ezin~\cite{BB}
generalized Wick theorem is not necessary in our case since we deal only
with Slater determinants).

The results of our projection (after variation) calculations for the
fission barrier of $^{240}$Pu for a positive
parity, are summarized in Figure~4.
\begin{figure}[h]
\begin{center}
\includegraphics[width=0.8\textwidth]{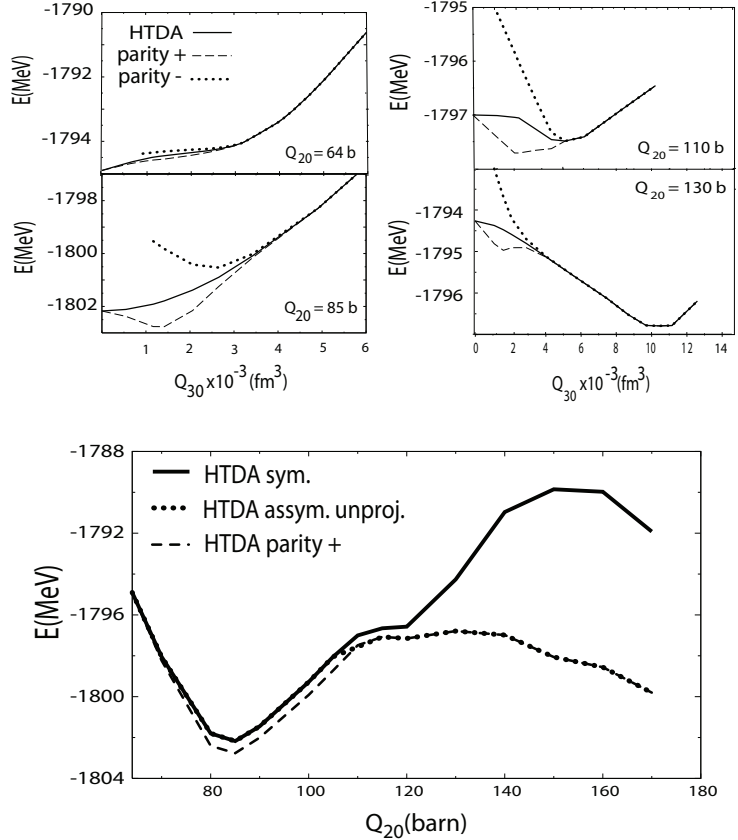}
\caption{Deformation energy curves for the $^{240}$Pu nucleus in
the fission isomeric state and second-barrier region. Upper panel:
energy curves as functions of the axial octupole moments for fixed values of
the axial quadrupole moment. Lower panel: resulting fission barriers.
}
\end{center}
\end{figure}
In the lower panel, it appears
that the projection has the following effects: i) it yields a
small stable octupole deformation around the fission isomeric state,
defined by an elongation $Q_{20}^{(0)}$, for an elongation much smaller
than the one ($Q_{20}^{(1)}$) where an instability of the HTDA unprojected
solution is observed, ii) before reaching the second (asymmetric)
fission barrier at an elongation $Q_{20}^{(2)}$, the projection does
not yield any significant effect on the energies.

This can be explained in the following way which is illustrated in the
upper panel of Fig. 4. Much before $Q_{20}^{(0)}$, at $Q_{20} =
64$~b for example, the unprojected solution is stable with respect to the axial
$Q_{30}$ mode with a stiffness large enough to only allow for the creation,
by the projection on positive parity, of a shoulder on the $E(Q_{30})$
curve at a fixed elongation. Around the superdeformed solution
$Q_{20}^{(0)} \sim 85$~b the stiffness parameter is no longer
sufficiently large to prevent the projection from creating a stable
equilibrium deformation (such a weak stiffness had already been found
in Hartree-Fock-Bogoliubov calculations in the super-deformation region of Hg and
Pb isotopes~\cite{SHBFM}).

At a quadrupole deformation around
$Q_{20} \sim 110$~b, the non-diagonal overlaps (for the identity and
Hamiltonian operators) become small enough so that the projected
energies are not significantly different from the unprojected ones, a
property which will be, of course, all the more verified than the
octupole deformation will increase. For even larger elongations,
e.g., $Q_{20} \sim 130$~b, the equilibrium octupole deformation
parameter will get larger to stabilize, as we have recalled, at a value
typical of the most probable asymmetric fission fragmentation and thus
yielding a minimum in the $E(Q_{30})$ curve exactly at the same place
than in the unprojected case.

As a result the height of the second fission barrier when projected on
the positive parity will be enhanced from its unprojected value. This
effect is only due to the relatively soft character of the octupole
deformation energy curve near the fission isomeric state and not to the
projection at the top of the second barrier where the octupole
deformation is too large to yield any projection effect.

It is our contention that these two properties (for the isomeric state
and at the top of the second fission barrier) are quite
general in the actinide region. Therefore, we deem that the underestimation of
such fission-barrier heights is probably a systematic effect. In our
current calculations, which are somewhat approximate, this enhancement
of the fission-barrier height is of about 350~keV.

%
%

\section{Conclusions}

We have discussed here two systematic effects leading to the
conclusion that usual microscopic calculations of actinide fission-barrier
heights could be underestimated by a few hundred keV for each.

One effect is related to the correlation existing between the quality of the
so-called Slater approximation for the Coulomb exchange energy and the
value of the proton single-particle level density. En passant, we note
that this defect (being related to a purely quantal property)
could not be possibly cured by a fit of EDF
parameters. Moreover it induces a
systematic lowering of the proton particle-hole energy which might
play a substantial role in the correlations built upon using such approximate
single-particle states.

The second effect has been established from parity-restoration
calculations from HTDA correlated solutions. Beyond the systematic
conclusion they seem to indicate, they correspond to the first
configuration-mixing calculations of such HTDA states where the
advantage of their Slater determinantal character has been fully
exploited.

\end{document}